\pdfoutput=1
\documentclass[11pt]{article}

\usepackage[utf8]{inputenc}
\usepackage[T1]{fontenc}
\usepackage{lmodern}
\usepackage[margin=1in]{geometry}
\usepackage{amsmath,amssymb}
\usepackage{graphicx}
\usepackage{booktabs}
\usepackage{xcolor}
\usepackage{algorithm}
\usepackage{algpseudocode}
\usepackage[numbers,sort&compress]{natbib}
\usepackage[colorlinks=true,linkcolor=blue,citecolor=blue,urlcolor=blue]{hyperref}
\usepackage{cleveref}

\title{Transpilation-Aware Multi-Vendor QPU Scheduling with\\
Predicted Post-Transpilation Fidelity}
\author{
  Ana Paula Appel\\
  Red Hat\\
  \texttt{aappel@redhat.com}
}
\date{November 2026}

\begin{document}
\maketitle

\begin{abstract}
Cluster schedulers already choose nodes, GPUs and accelerators for submitted workloads, enforce quotas and
spread load. QPUs from several vendors are now schedulable resources in the same clusters, yet the user
still names the device, usually the one with the best average fidelity. For a given circuit that choice
is often wrong: compiling onto each device's qubit connectivity adds gates, so a job can meet its
required success probability on one device and miss it on another with similar error rates, and checking
this exactly by compiling for every candidate is too slow for a scheduling decision.

We let the scheduler choose the QPU in two stages. A training-free estimate predicts, in milliseconds and
without compiling, the success probability on every candidate and filters out those that would miss the
requirement; only the chosen device then compiles the circuit, re-checks it, and binds the job or rolls
back. The estimate matches compiling every candidate at about $25\times$ lower latency in simulation,
ranks measured success on IBM hardware as well as the post-compilation estimate ($\rho=0.95$), reduces
requirement violations from tens of jobs to a few, and avoids them under contention without piling jobs
onto one device. Across vendors it works when error rates are comparable (IBM and IQM) and fails when
published figures overstate a device.
\end{abstract}

\section{Introduction}\label{sec:intro}
Quantum backends now sit behind the same cluster managers that already place CPUs and GPUs. HPC sites
expose them through Slurm hierarchical resources and interfaces such as QDMI~\cite{mqsc2026} and
QRMI~\cite{qrmi2026,ibmfermilab2026}; cloud and container platforms do the same with Kubernetes operators
and custom resources~\cite{qonductor2024,qubernetes2024}. Those layers allocate and account for QPU time.
They do not, as a rule, decide \emph{which} device should run a given circuit.

In a production cluster this is the same decision the scheduler already makes for every other
resource. A user submits a workload (for example a variational optimisation loop or a batch of sampling
jobs inside a container) and states what it needs; the scheduler picks the node, the GPU and the
accelerator, enforces quotas and spreads load, without the user naming a machine. A QPU could be chosen
the same way: the job declares its circuit, a required success probability and a budget, and the
scheduler places it on a device that meets the requirement at the lowest cost or the shortest wait.
Automatic choice matters more as the number of devices and vendors grows, as calibration changes during
the day, and as many users share the same pool, because a manual choice is fixed when the job is written
and cannot react to queues or quotas at submission time.

That decision is still made by the user, and it is easy to get wrong for three reasons.
\begin{itemize}
  \item \textbf{Admission ignores circuit quality.} Current integrations treat a QPU as available once
        the status probe succeeds. The job then runs on the device named in the request, regardless of
        calibration, price or queue. A job can consume paid QPU time and still miss its fidelity floor.
  \item \textbf{Average device fidelity is the wrong signal.} Routing inserts SWAPs whenever the circuit
        interaction graph does not embed in the coupling map, and two devices with similar mean error
        rates can produce very different compiled gate counts for the same circuit. In our measurements,
        picking the backend with the best average fidelity put 42--74 of 102 jobs below their requested
        success probability.
  \item \textbf{The exact check does not fit the hot path.} One level-1 transpilation already takes
        5\,ms--3\,s for circuits of 10--80 qubits. Doing that for every candidate inside the filter stage
        does not scale with the number of backends.
\end{itemize}

We therefore split the decision into two stages: a millisecond estimate over all candidates, and an
exact transpile-and-check only for the reserved one.

\paragraph{Contributions.} Experiments are labelled E1--E12 and detailed in
\cref{sec:setup,sec:results}. The prototype, experiment scripts and results are available at
\url{https://github.com/anapaulaappel/QPUScheduler}.
\begin{enumerate}
  \item A training-free topology-divergence estimate of post-routing two-qubit overhead, with Spearman
        $\rho\approx0.99$ against actual gate counts at millisecond cost (\cref{sec:method}; E1, E1b).
  \item A two-stage protocol that uses the resulting estimated success probability as a hard filter
        (estimate, reserve, transpile, re-validate, bind or roll back), expressed in generic scheduler
        hooks and mapped to the Kubernetes Scheduling Framework~\cite{k8sframework} and to Slurm
        (\cref{sec:method}).
  \item Simulation evidence that selection by the estimate approaches transpile-all on regret and SLA
        safety at about $25\times$ lower latency (E2, E3), behaviour under contention among jobs with cost
        and quota accounting (E11), and negative results for per-qubit and routing-replay variants.
  \item Hardware measurements on six devices from four vendors: mirror-circuit ranking and selection
        (E6, E7), application circuits (E10), policy replay with conformal margins (E9), a comparison
        against a learned predictor (E8), a rerun under four-day-old calibration, a controlled same-day
        cross-vendor run (E12), and a layout-matched diagnostic; 95\% bootstrap intervals for the main
        claims.
  \item A cross-pricing usage model (\cref{sec:method}) and a negative result on forecast-based deferral
        over 11.6 days of real IBM calibration history (E4, E5).
\end{enumerate}

\section{Related Work}\label{sec:related}
\paragraph{Cluster integration of QPUs.} Qonductor~\cite{qonductor2024} is a Kubernetes-based
quantum cloud orchestrator whose hybrid scheduler balances fidelity and completion time across
IBM devices. Qubernetes~\cite{qubernetes2024} and Kueue/Argo-based
frameworks~\cite{kueuequantum2026} run quantum tasks as Kubernetes jobs. SQUIRO~\cite{squiro2026}
selects among IBM and IonQ backends on Kubernetes with a hard feasibility mask (qubits,
capabilities, calibration freshness) followed by a weighted score with a topology-mismatch
penalty, without modelling transpilation. Fluence~\cite{fluence2026} is a Kubernetes scheduler
plugin that chooses the cheapest or shortest-queue Braket backend. On the HPC side, the Orchestrator of the Munich
Quantum Software Company (MQSC)~\cite{mqsc2026} integrates QPUs into Slurm through hierarchical resources (HRES) and
the QDMI device interface; a device is admitted when its status probe reports it available, and
the job compiles for and runs on the device the user requested. IBM's QRMI~\cite{qrmi2026},
announced with Fermilab and NVIDIA~\cite{ibmfermilab2026}, is a vendor-agnostic layer that also
plugs into Slurm HRES to allocate, reserve and account for QPU time. These layers provide
allocation, admission and accounting; none selects a device by the circuit's predicted fidelity
after transpilation. Our method decides which device to request and can be hosted by any
of them (\cref{sec:integration}).

\paragraph{Prediction before compilation.} The closest work is Fulginiti et
al.~\cite{fulginiti2026}, who predict with a graph neural network, before compilation, the
expected fidelity of each circuit on each QPU of an HPC system and compile only for the device
the scheduler chooses. Their devices are emulated partitions of one hardware family, and the
scheduler trades predicted fidelity against parallelism as a weighted score. MQT Predictor~\cite{mqtpredictor}, from the Munich Quantum Toolkit (MQT), selects the best of seven superconducting and trapped-ion devices
with a classifier, without compiling, and then compiles for that device; Salm et
al.~\cite{salm2023} recommend a compiler--QPU pair before compilation from past executions.
Q-fid~\cite{qfid2023}, Hartnett et al.~\cite{hartnett2024} and Das et al.~\cite{das2025gnn}
learn fidelity or error rates from circuit and device features. Predicting before compiling is
therefore not our contribution. We add a training-free estimate whose inputs are the device's
calibration and coupling map, its use as a hard constraint with exact re-validation and rollback,
validation of ranking and selection on measured hardware from four vendors, and a direct
comparison with a learned predictor on unseen devices (E8).

\paragraph{Quality- and cost-aware scheduling.} Qurator~\cite{qurator2026} schedules hybrid
workflows across six providers with a unified success score from calibration data and a
user-bounded fidelity loss, without the circuit's transpilation. QOS~\cite{qos2025} and Ravi et
al.~\cite{ravi2021} estimate fidelity across IBM devices and trade it against waiting time.
DRLQ~\cite{drlq2024} and QFOR~\cite{qfor2025} place tasks with reinforcement learning on current
state. Quantum Broker~\cite{quantumbroker2026} splits shots across providers under declared cost,
latency and fidelity constraints, and Shinde and Szefer~\cite{shinde2026} show that
fidelity-per-cost rankings differ from fidelity rankings.

\paragraph{Compilation, calibration and benchmarking.} The SABRE routing algorithm~\cite{sabre2019},
Qiskit~\cite{qiskit2024} and tket~\cite{tket2020} perform placement and routing for a given
target; error-aware compilation~\cite{nishio2020} uses the estimated success probability (ESP)
to choose qubits. Rodr\'iguez-Soriano et al.~\cite{rodriguezsoriano2025} forecast two-qubit error
rates to compensate queue delay at compile time, and Kurniawan et al.~\cite{kurniawan2024} study
calibration data in error-aware compilation; we tried forecasts for deferral instead (E4, E5).
Our labels follow mirror-circuit~\cite{proctor2022} and application-oriented
benchmarking~\cite{lubinski2021} practice.

\section{Problem Formulation}\label{sec:background}
A circuit $c$ has interaction graph $G_c=(V_c,E_c,w)$, where $w(u,v)$ counts two-qubit gates on
$(u,v)$. A QPU $q$ has coupling graph $G_q$ and native gate set $B_q$. Transpilation finds a
placement $\phi:V_c\to V_q$, inserts SWAPs so that every two-qubit gate acts on adjacent qubits,
and rewrites gates into $B_q$. A job $J=(c,\tau,C_{\max},\text{deadline})$ requires that the
fidelity $F(c,q)$ of the transpiled circuit satisfies $F(c,q)\ge\tau$ within cost and quota limits.
$F(c,q)$ must be predicted for every candidate within the scheduler's latency budget.
We use three measures throughout. A violation of the service-level agreement (\emph{SLA violation}) is a job placed on a device where its
true (simulated or measured) success is below $\tau$. A \emph{false rejection} is a job declared
unschedulable although some candidate met $\tau$. The \emph{regret} measures how much success is lost by not
choosing the best device: for each circuit it is the true success on the best candidate minus the true
success on the chosen one, averaged over circuits. For example, if a circuit would succeed with
probability 0.80 on the best device and the scheduler chooses one where it succeeds with 0.78, the
regret for that circuit is 0.02; a mean regret of 0.008 means that, on average, 0.8 percentage points of
success are lost relative to a perfect choice. The \emph{top-1} rate is the share of circuits for which
the chosen candidate is the best one (ties count as hits). The two are complementary: top-1 counts how
often the choice is exactly right, regardless of how close the alternatives are, while regret measures
how much a wrong choice costs.

\section{Method}\label{sec:method}
\paragraph{QPU state.} A read-only telemetry agent writes, per backend, error rates, $T_1/T_2$,
coupling map, native gates, operation durations, pricing, queue, calibration age and a quality
forecast into platform resource records; stale telemetry excludes the backend.

\paragraph{Topology-divergence estimate.} (i) Build $G_c$ in a neutral $\{u,\mathrm{cx}\}$ basis;
(ii) compute distances $d_q$ on $G_q$; (iii) place by subgraph embedding with the VF2 algorithm~\cite{vf2} when one
exists (SWAP-free), otherwise greedily, keeping strongly interacting qubits close; (iv) estimate
\begin{equation}
  S=\sum_{(u,v)\in E_c} w(u,v)\,\max\bigl(0,\,d_q(\phi(u),\phi(v))-1\bigr),\qquad
  \Delta N_{2q}=k_{\mathrm{swap}}(q)\,S,
\end{equation}
with $k_{\mathrm{swap}}=3$ on sparse devices and $0$ on all-to-all devices; (v) estimate a
gate-set translation overhead $r_{\mathrm{gate}}$ from per-native-gate rules. A variant replays the
two-qubit gates with a greedy router (``routing replay''). The raw estimate is weaker than SABRE; a
single factor $\alpha\approx0.28$, fitted leave-one-family-out (LOFO), calibrates its magnitude.

\paragraph{Predicted fidelity as a hard filter.} The predicted fidelity is an estimated success
probability (ESP): the probability that a run of the transpiled circuit has no error, from the
device-mean error rates and the estimated gate counts. In the tables, ``ESP, topology estimate'' is
$\widehat{\mathrm{ESP}}$ below, and ``ESP, transpiled circuit'' uses the actual counts after transpilation.
\begin{equation}
  \widehat{\mathrm{ESP}}(c,q)=(1-\bar e_1)^{r_{\mathrm{gate}}N_1}\,(1-\bar e_2)^{N_2+\alpha\Delta N_{2q}}\,
  (1-\bar e_{\mathrm{ro}})^{n}\,e^{-\hat t/T_2},
\end{equation}
with device-mean error rates $\bar e$ and estimated duration $\hat t$. A candidate passes if
$\widehat{\mathrm{ESP}}-m\ge\tau$ for a margin $m\ge0$. A learned model can replace the estimate if
its validation metrics meet an operator-defined acceptance threshold; otherwise the physics
estimate is used.

\begin{figure}[t]
  \centering
  \includegraphics[width=\linewidth]{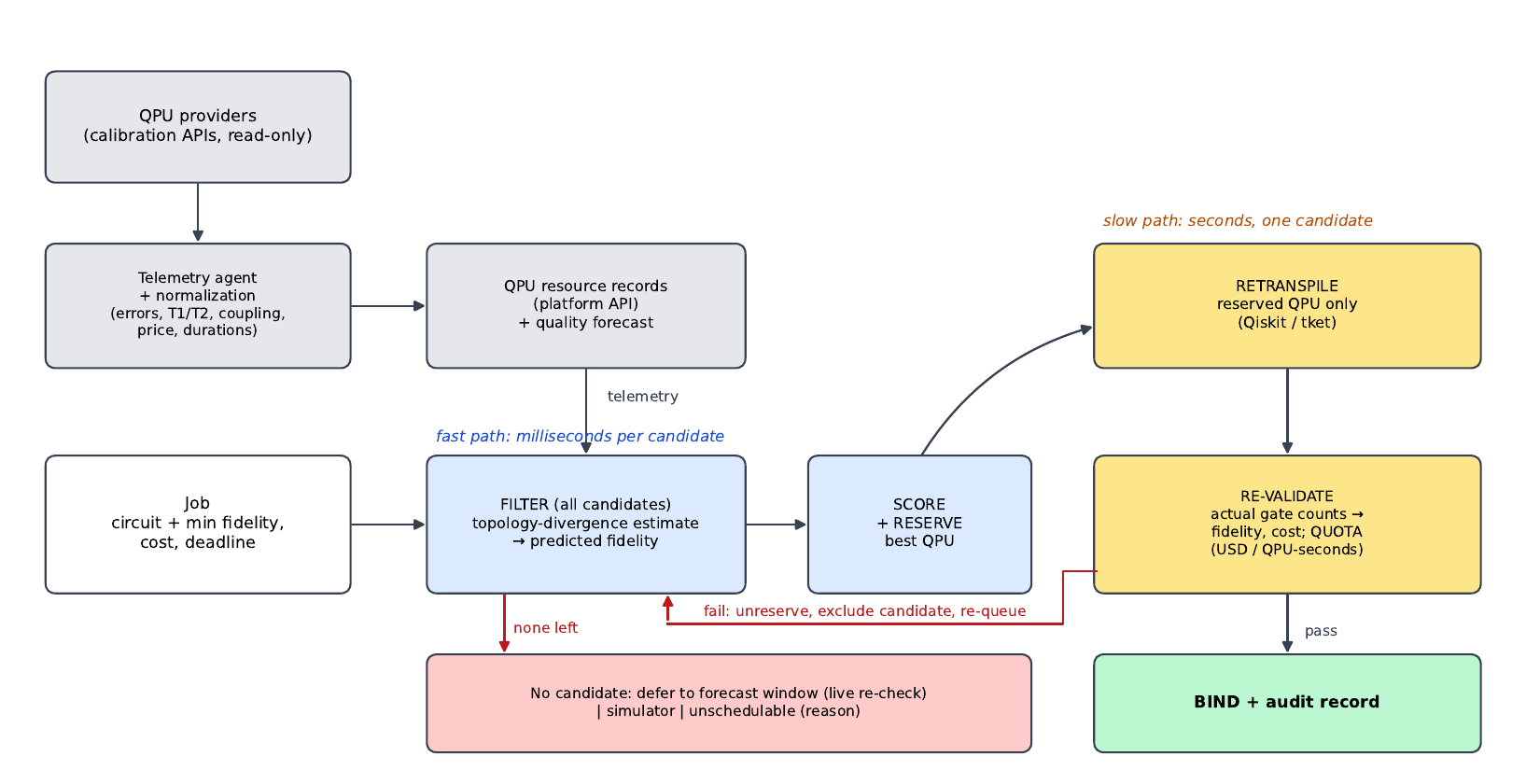}
  \caption{Two-stage, transpilation-aware placement. Provider telemetry is normalized into per-QPU
resource records. Given a job (circuit, fidelity floor, cost limit, deadline), the fast stage scores
every candidate with the topology-divergence estimate and reserves the best one that passes the filter;
the slow stage transpiles only for that QPU, re-checks fidelity and cost from the actual gate counts,
charges the quota, and either binds the job or releases the reservation and re-queues. If nothing
qualifies, the job is deferred, sent to a simulator, or rejected with a reason.}
  \label{fig:arch}
\end{figure}

\paragraph{Two-stage protocol.} \Cref{fig:arch} shows the pipeline. Candidates are filtered with
the estimate; the best is scored and reserved; the circuit is transpiled for the reserved QPU only
(Qiskit, or tket for trapped-ion targets) and re-validated with the actual gate counts
(fidelity and cost); usage is reserved against quotas; the job is bound, or the reservation is
released, the candidate excluded and the job re-queued, up to a retry limit. When no candidate
qualifies, the job is deferred to a forecast quality window (with a live re-check), sent to a
simulator, or marked unschedulable with a reason.

\paragraph{Usage and quota.} QPU time is estimated as
$t=\text{shots}\cdot(\text{circuit duration}+\text{reset})+\text{overhead}$. Pay-per-use cost
combines task, shot, gate-shot and per-second terms; for prepaid QPU-time plans the effective cost
is $t\cdot(\text{plan price}/\text{plan time})$. Quotas are enforced in native units (USD, QPU
seconds, shots) with optimistic concurrency, and scoring uses the effective cost.

\paragraph{Quality windows.} A per-backend forecast with an uncertainty band (calibration-cycle
model or Prophet~\cite{prophet2018}) defines a quality window as a maximal run of steps whose
lower bound is at least $\tau$; deferred jobs are fully re-scheduled at the window start.

\paragraph{Integration with cluster schedulers.}\label{sec:integration} The protocol needs four
hooks from the host scheduler: a per-candidate filter that can reject a device for a given job, a
score, a reservation that can be released, and a step between reservation and binding where the
job can be modified (transpiled) and re-checked. \Cref{tab:integration} maps them onto two hosts.
The prototype implements the phases as a stand-alone pipeline over an internal candidate set and
is not deployed as a plugin of either scheduler. In Slurm the user names the device in the
request, so the natural insertion point is a broker that chooses the device before submission,
with the prolog performing the exact transpilation and re-validation.

\begin{table}[t]
\centering\small
\caption{How the four hooks the protocol needs (filter, score, releasable reservation, and a step between
reservation and binding) map onto the Kubernetes Scheduling Framework and onto an HPC workload manager
(Slurm). The prototype implements these phases as a stand-alone pipeline.}\label{tab:integration}
\resizebox{\linewidth}{!}{\begin{tabular}{@{}lll@{}}
\toprule
Phase & Kubernetes Scheduling Framework & HPC workload manager (Slurm)\\
\midrule
Candidate set & QPUs as custom resources; PreFilter & QPUs as HRES (QDMI/QRMI) or licenses\\
Filter (estimate) & Filter plugin & submit/select plugin or pre-submission broker\\
Score & Score plugin & priority or site factor\\
Reserve, transpile, re-validate & Reserve, Permit, PreBind & prolog inside the allocation\\
Rollback & Unreserve, re-queue & requeue with the device excluded\\
Defer to a window & scheduling gate & \texttt{-{}-begin} time or job hold\\
\bottomrule
\end{tabular}}
\end{table}

\section{Experimental Setup}\label{sec:setup}
\paragraph{Targets.} Five heavy-hex devices (Fez, Torino, Brisbane, Sherbrooke, Kyiv) and one
square-lattice device (Nighthawk), using IBM fake backends that carry real calibration snapshots;
two trapped-ion-like all-to-all devices (native two-qubit gate $R_{XX}$; single-qubit (1Q) error $2\cdot10^{-4}$, two-qubit (2Q) error
$5\cdot10^{-3}$, readout $4\cdot10^{-3}$, jittered per qubit).

\paragraph{Circuits.} Seven families: Greenberger--Horne--Zeilinger (GHZ) states, the quantum Fourier transform (QFT), random circuits, the quantum approximate optimization algorithm (QAOA) on a ring and on a complete graph,
RealAmplitudes with full entanglement, and Bernstein--Vazirani. E1 uses 35 circuits of 4--12 qubits (280 placements on the eight targets). E2, E3 and the
hardware validation use 102 circuits of 4--14 qubits, with three random instances per family and
size where the family allows it (816 simulations on the eight targets for E2).

\paragraph{Why these circuits.} The benchmark is designed for experimental control. The families
cover different interaction patterns, from a linear chain (GHZ) to all-to-all connectivity (QAOA on a
complete graph), so they stress the topology-divergence estimate in different ways: the same circuit
pays a different routing cost on a heavy-hex, a square-lattice or an all-to-all device. Sizes and random
instances are fixed with seeds, three instances per family and size, so the effect of topology can be
separated from the effect of size and from the luck of a single instance, and every experiment can be
reproduced. The same circuits are used in E2, E3, E6, E7, E9 and the rerun with older calibration, so
comparisons are paired circuit by circuit. Finally, mirror circuits (below) give each circuit a single
known ideal output, which makes the success probability well defined and comparable across devices,
vendors and qubit technologies, without simulating the ideal output distribution. E10 complements them
with application circuits, to check how far the results carry to realistic workloads.

\paragraph{Labels.} We use mirror circuits~\cite{proctor2022}: a random $X$ layer, $C$, a barrier,
and $C^\dagger$. The random $X$ layer makes the ideal output a random bitstring rather than
$|0\ldots0\rangle$, so that readout errors and a bias towards the all-zero state cannot pass as success. The ideal output is a single bitstring, and the label is its success probability
in a noisy Aer simulation of the circuit transpiled for the target, using the target's noise model.
Raw Hellinger fidelity of the plain circuits is unusable as a label: QFT$|0\ldots0\rangle$ has a
uniform ideal output, and noise pushes near-uniform outputs towards uniform without lowering the
fidelity. In noiseless simulation every mirror circuit returns its ideal bitstring with probability
one. Actual transpilation uses Qiskit at optimization level 1, best of three seeds.

\paragraph{Hardware (E6).} The 102 mirror circuits of E2 were run on \texttt{ibm\_fez},
\texttt{ibm\_marrakesh} and \texttt{ibm\_kingston} (IBM Heron r2), 2000 shots each, transpiled for
each device (306 executions, 177\,s of QPU time on the free Open Plan; our usage model predicted
178\,s). The label is the measured success probability. An earlier run used the 28 circuits of
4--10 qubits with 4000 shots.

\paragraph{Other vendors (E7).} The 28 circuits of the earlier E6 run (4--10 qubits) ran on IQM Garnet (20 qubits, square
lattice, native $\{r, \mathrm{cz}\}$; IQM Resonance free plan, 500 shots, error rates from the
provider's calibration metrics of the day) and 12 of them on Rigetti Cepheus-1-108Q (native
$\{r_z, r_x(\pm\pi/2), r_x(\pi), \mathrm{cz}\}$; Open Quantum free credits, 1000 shots). For Rigetti
no job calibration was available, so we use the published median errors (single-qubit $0.1\%$,
controlled-Z (CZ) $0.9\%$) and assume a $3\%$ readout error. The Rigetti circuits are submitted inside a verbatim
box with physical qubits: without it, the provider's compiler optimized across the barrier of the
mirror circuits, and a circuit with 256 CZ gates returned its ideal bitstring with probability
$0.74$ (with the verbatim box, $0.005$). Mirror-circuit benchmarks on cloud devices therefore need
compilation to be disabled. Six of the circuits (one per family at 6 qubits, without the complete-graph
QAOA) also ran on the AQT IBEX Q1 trapped-ion device (12 qubits, all-to-all, native
$\{\mathrm{prx}, r_z, \mathrm{xx}\}$; Open Quantum free credits, 100 shots, so a standard error of up
to $0.05$), in a verbatim box as well. We use AQT's published figures (single-qubit fidelity
$99.97\%$, two-qubit $98.85\%$) and assume a $0.5\%$ readout error. A local statevector simulation
of every verbatim file returns the ideal bitstring with probability one.

\paragraph{Controlled cross-vendor run (E12).} E7 mixes days and shot counts. To control both, on
20 Nov 2026 we ran the 28 circuits of the earlier E6 run (same seeds, hence identical circuits) on the
three IBM Heron devices and on IQM Garnet, and the 12 circuits with 4 and 8 qubits also on Rigetti
Cepheus-1-108Q (verbatim), all with 1000 shots on the same day. IBM uses the calibration read at
preparation, IQM the error rates of E7 and Rigetti the published medians. AQT was not included: 12
circuits at 1000 shots exceeded the available credits.

\paragraph{Code and data.} All scripts, measured results, the error rates used for the non-IBM devices and the
figures of this paper are in the project repository\footnote{\url{https://github.com/anapaulaappel/QPUScheduler}, Apache-2.0.};
each experiment is one script under \texttt{experiments/} that writes its results to
\texttt{experiments/results/}. Historical IBM calibration can be retrieved by date through the IBM
Quantum API.

\paragraph{Uncertainty.} Confidence intervals are 95\% percentile intervals from a paired bootstrap over
circuits ($B=2000$, fixed seed): each resample draws circuits with replacement and keeps all devices of a
drawn circuit, because the observations of one circuit on several devices are not independent
(\cref{app:ci}).

\section{Results}\label{sec:results}

\begin{figure}[t]
  \centering
  \includegraphics[width=0.8\linewidth]{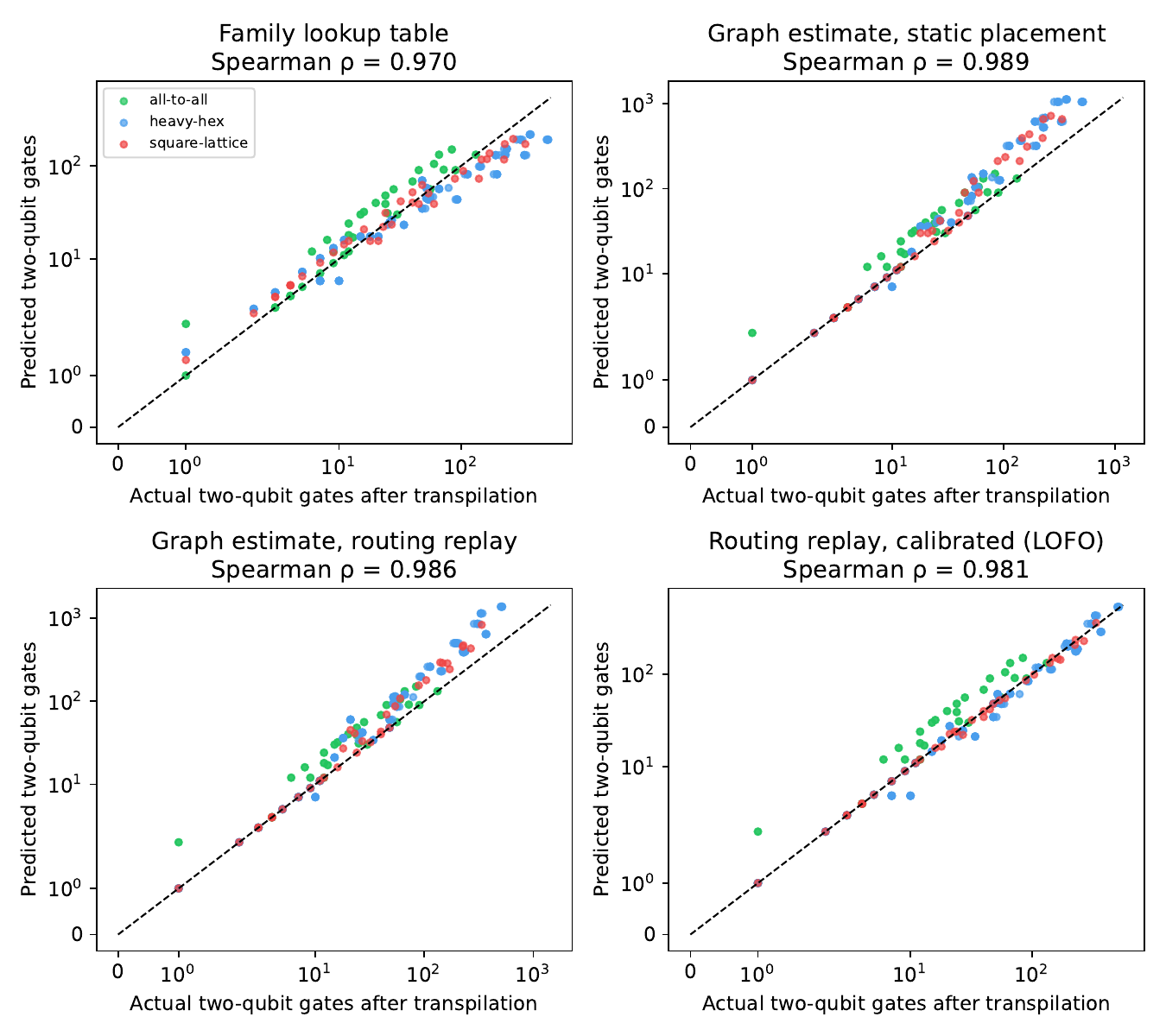}
  \caption{E1: predicted (vertical) vs.\ actual (horizontal) two-qubit gates after transpilation, log--log,
one point per (circuit, target), coloured by topology. The dashed line is perfect prediction. Panels
compare a per-family lookup table, the graph estimate with static placement, the same estimate with
routing replay, and routing replay after the LOFO calibration factor. The graph-based predictors follow
the diagonal across heavy-hex, square-lattice and all-to-all targets; the lookup table does not.}
  \label{fig:e1}
\end{figure}

\paragraph{E1: estimate vs.\ actual overhead.} \Cref{tab:e1} and \cref{fig:e1} compare predicted and
actual two-qubit gate counts over 280 placements, by Spearman $\rho$, mean absolute error (MAE) and median relative error. The graph estimate ranks best; one calibration
factor fixes its magnitude.

\begin{table}[t]
\centering
\caption{E1: how well each predictor estimates the number of two-qubit gates after transpilation, over 280
(circuit, target) placements on eight simulated devices. Spearman $\rho$ measures ranking, MAE the mean
absolute error in gates, and the last column the median relative error. The graph estimate ranks best;
one calibration factor, fitted leave-one-family-out (LOFO), fixes its magnitude.}
\label{tab:e1}
\begin{tabular}{lccc}
\toprule
Predictor & Spearman $\rho$ & MAE & Median rel.\ error\\
\midrule
Ignore transpilation overhead & 0.946 & 40.9 & 41\%\\
Family lookup table & 0.970 & 30.5 & 31\%\\
Graph estimate, static (raw) & \textbf{0.989} & 93.1 & 51\%\\
Graph estimate, static (calibrated, LOFO) & 0.979 & 20.9 & 12\%\\
Graph estimate, routing replay (calibrated, LOFO) & 0.981 & \textbf{12.4} & \textbf{7\%}\\
\bottomrule
\end{tabular}
\end{table}

\begin{figure}[t]
  \centering
  \includegraphics[width=0.7\linewidth]{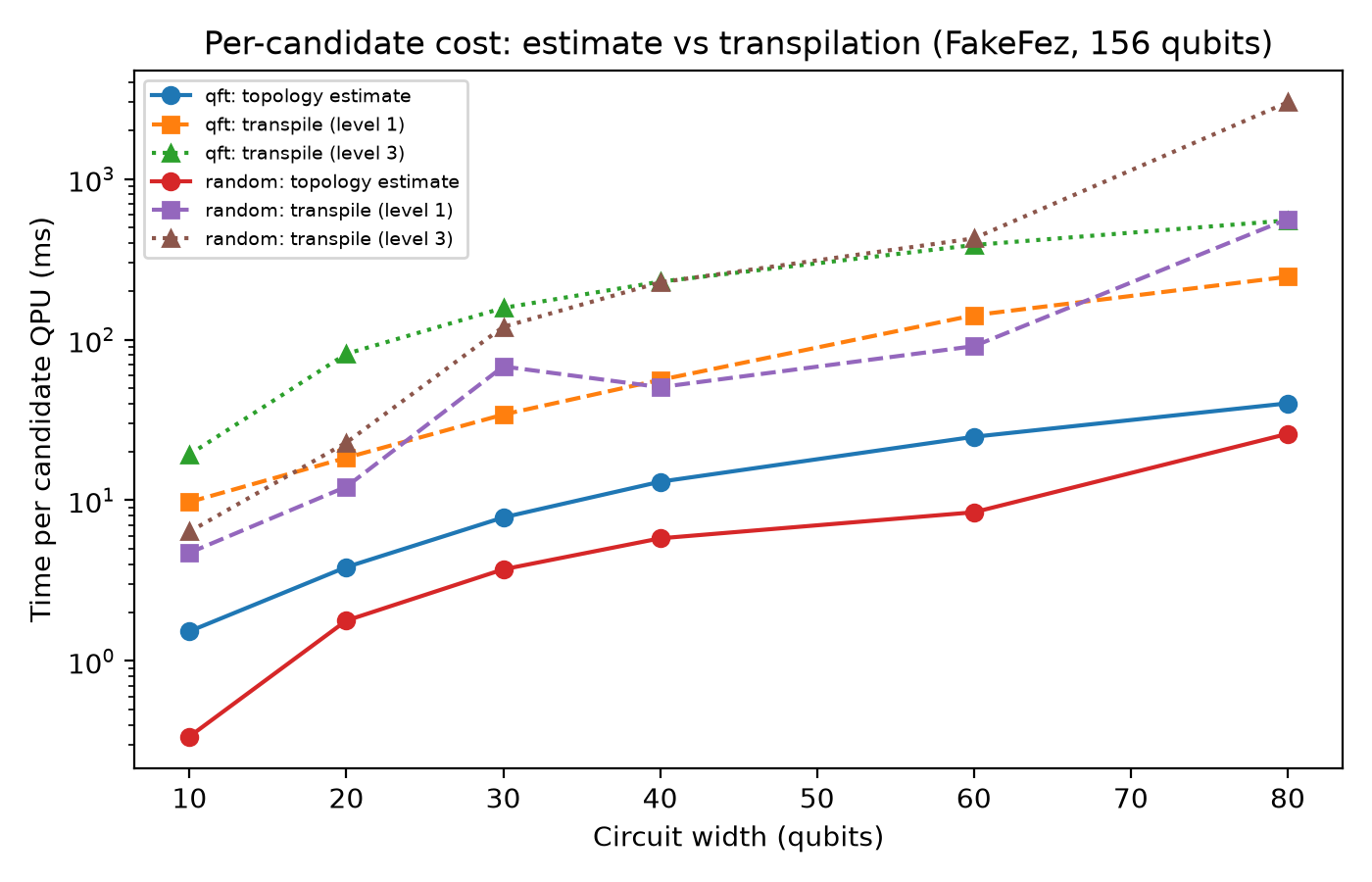}
  \caption{E1b: wall-clock time to score one candidate on a 156-qubit heavy-hex device, as a function of
circuit width, for the topology estimate and for Qiskit at optimization levels 1 and 3 (curves split by
circuit family). The estimate is about $4$--$22\times$ faster than level~1, which is the budget that lets
the filter evaluate every backend instead of compiling for each.}
  \label{fig:e1b}
\end{figure}

\paragraph{E1b: latency.} On a 156-qubit heavy-hex device, the estimate is about $4$--$22\times$ faster
than a level-1 transpilation and $13$--$115\times$ faster than level 3. At 80 qubits that is
26--40\,ms vs.\ 0.25--0.56\,s and 0.55--3.0\,s (\cref{fig:e1b}); timings vary by about $\pm30\%$ between runs.

\begin{figure}[t]
  \centering
  \includegraphics[width=\linewidth]{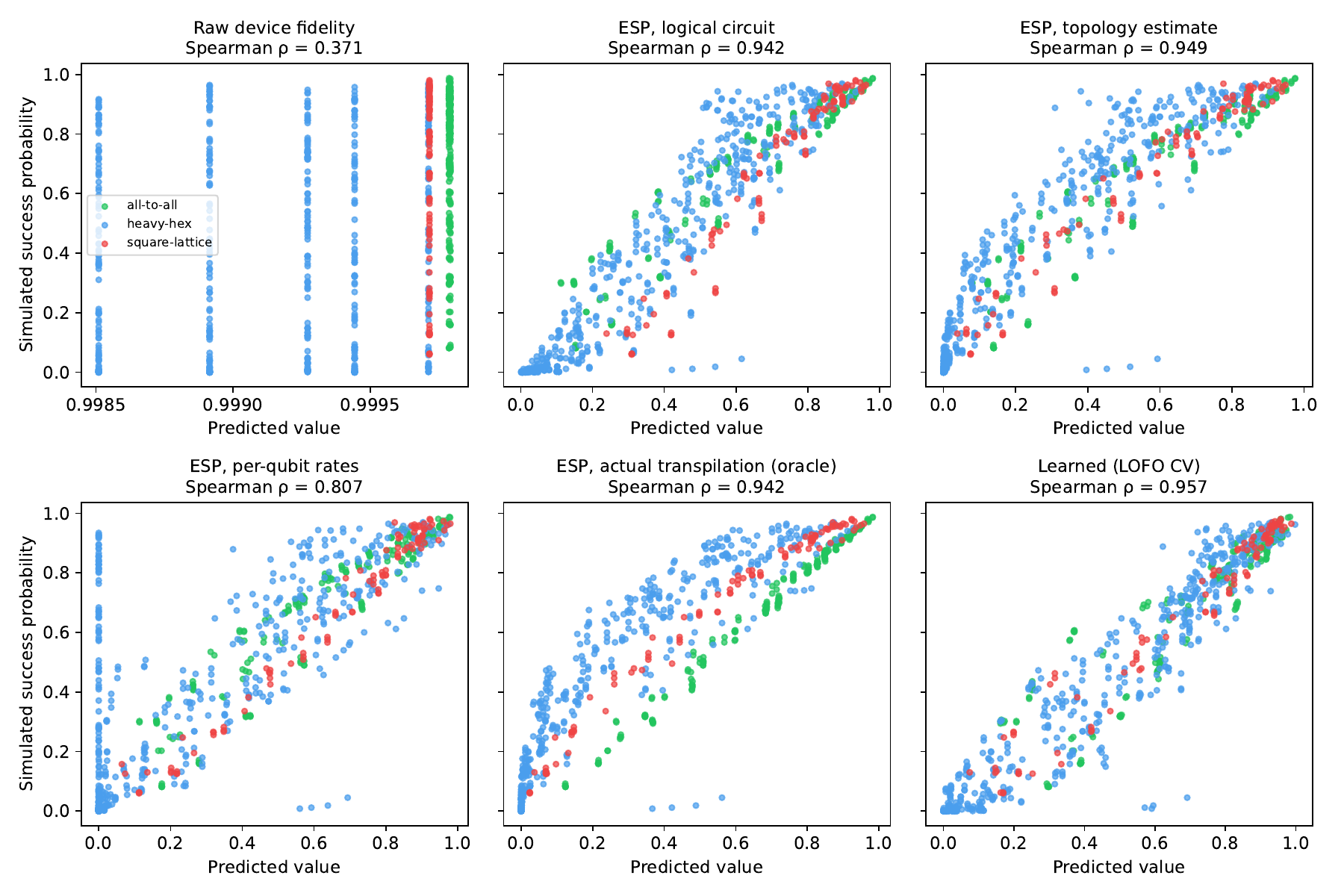}
  \caption{E2: each predictor (horizontal) against simulated success probability (vertical), one point per
(circuit, device) pair, coloured by topology. Raw device fidelity ignores the circuit and scatters;
ESP from the topology estimate tracks the label about as tightly as ESP computed from the actually
transpiled gate counts.}
  \label{fig:e2}
\end{figure}

\paragraph{E2: predicting success without transpiling.} Across 816 simulations of 102 circuits
(\cref{tab:e2}, \cref{fig:e2}), the success probability estimated without transpiling ranks
circuits as well as the one computed from the transpiled circuit, and choosing the target with it
comes close to transpiling every candidate: its mean regret is $0.016$ against $0.015$, but it
picks the best target less often (32\% vs.\ 41\%), so the remaining gap is in near-ties. In an
earlier run with 28 circuits both had the same regret ($0.0129$). Raw device fidelity is a poor
predictor. Two richer variants did worse at selection: the routing replay ranks slightly better
but picks worse targets, and per-qubit error rates are worse on both counts, probably because the
compiler's actual layout differs from our placement.

\begin{table}[t]
\centering
\caption{E2: predictors of the simulated success probability of 102 mirror circuits on eight devices (816
simulations). Spearman $\rho$ measures how well each predictor ranks circuits; the mean regret is the
success lost, on average, by choosing the device the predictor ranks first instead of the best one. The
estimate without transpilation ranks as well as the estimate from the transpiled circuit and selects
targets almost as well.}
\label{tab:e2}
\begin{tabular}{lcc}
\toprule
Predictor & Spearman $\rho$ & Mean regret of target choice\\
\midrule
Raw device fidelity & 0.37 & 0.019\\
ESP, logical circuit & 0.94 & 0.041\\
\textbf{ESP, topology estimate (no transpilation)} & \textbf{0.95} & 0.016\\
ESP, routing-replay estimate & 0.95 & 0.026\\
ESP, per-qubit errors & 0.81 & 0.036\\
ESP, actual transpilation (oracle) & 0.94 & 0.015\\
Learned (leave-one-family-out) & 0.96 & 0.014\\
\bottomrule
\end{tabular}
\end{table}

\begin{figure}[t]
  \centering
  \includegraphics[width=\linewidth]{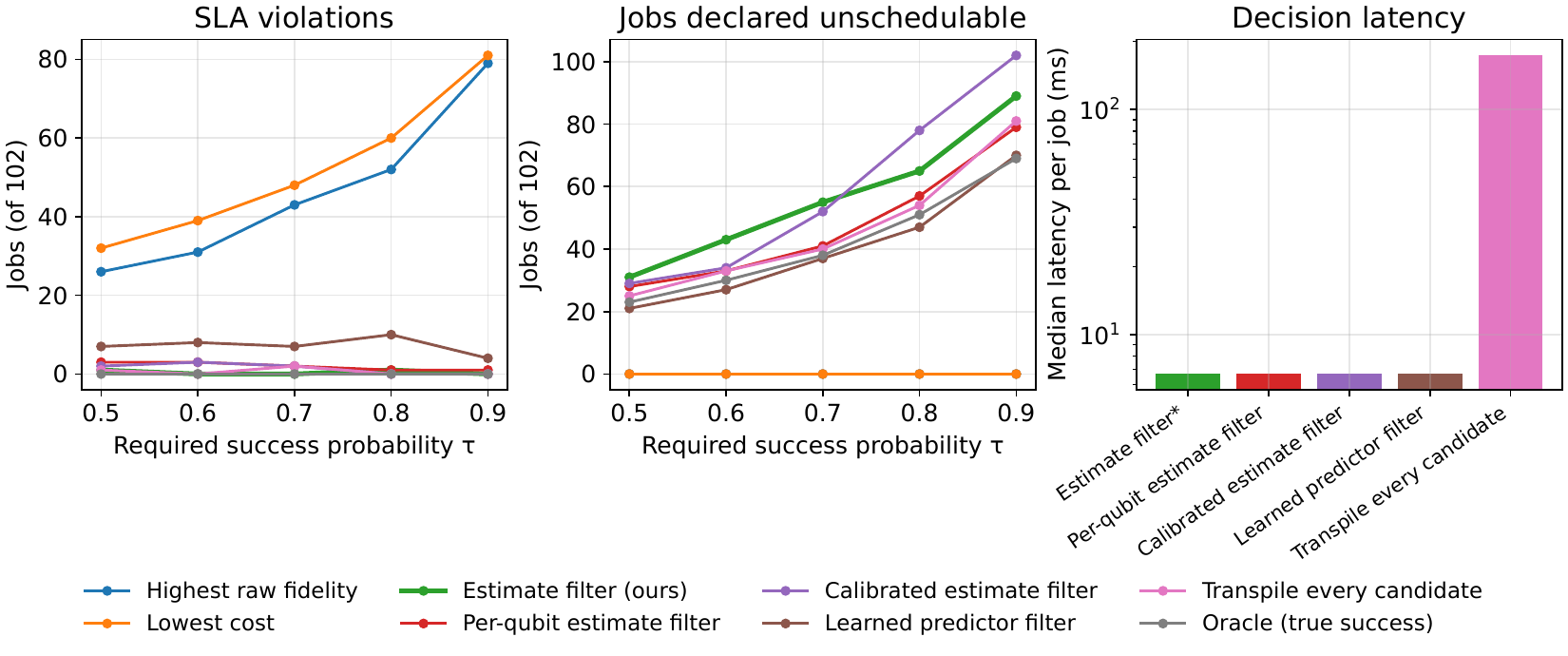}
  \caption{E3: placement policies on 102 jobs and eight simulated candidates as the required success
probability $\tau$ varies. Left: SLA violations (jobs that finished below $\tau$). Middle: jobs declared
unschedulable. Right: median decision latency per job. Filtering with the estimate keeps violations near
zero, like transpile-all, at a fraction of the latency; raw-fidelity and lowest-cost placement miss the
floor for many jobs.}
  \label{fig:e3}
\end{figure}

\paragraph{E3: scheduling policies.} With the 102 circuits of E2 as jobs and 8 candidates
(\cref{fig:e3}), placing on the highest raw fidelity violates the requirement in 31/52/79 jobs at
$\tau=0.6/0.8/0.9$, and lowest-cost placement in 39/60/81. The estimate filter has 0/1/0
violations, with 13/14/20 false rejections, at about 7\,ms per job. Transpile-all has 0/0/0
violations, with 3/3/12 false rejections, at about 174\,ms. Over all thresholds from 0.5 to 0.9 the
estimate filter has at most one violation per threshold and transpile-all at most two, since both
predict from device-mean error rates. In an earlier run with 28 jobs neither had violations.
Isotonic calibration traded fewer false rejections at low $\tau$ for 2--3 violations, and more
rejections at high $\tau$; a learned filter rejects less but allows 4--10 violations.

\begin{figure}[t]
  \centering
  \includegraphics[width=\linewidth]{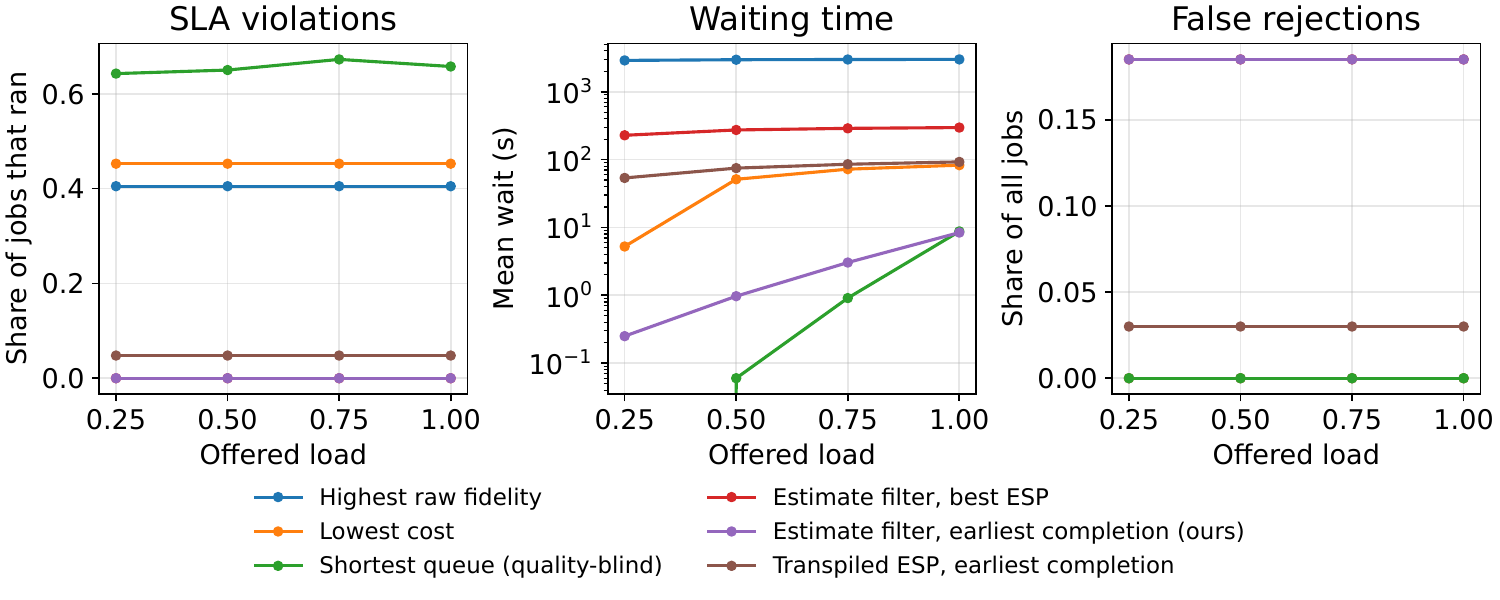}
  \caption{E11: 400 jobs with Poisson arrivals competing for eight simulated devices, as a function of
the offered load (relative to the capacity of each job's fastest device; $\tau=0.7$). Left: share of
the jobs that ran and finished below $\tau$. Middle: mean waiting time (log scale). Right: share of jobs
rejected although some device met $\tau$. Choosing the device with the best average fidelity piles all
jobs onto one QPU and still violates the requirement; a quality-blind balancer violates it for most jobs;
the estimate filter with earliest-completion placement keeps violations at zero and waiting times short,
at the cost of rejecting some feasible jobs.}
  \label{fig:x1}
\end{figure}

\paragraph{E11: contention among jobs.} E3 places each job in isolation. To test the filter under load,
a discrete-event simulation sends 400 jobs (E2 circuits drawn with replacement, $\tau=0.7$) with Poisson
arrivals to the eight simulated devices, each with a first-in-first-out queue; service time is the
QPU time of the usage model on the transpiled circuit, and the offered load is measured against the
capacity of each job's fastest device (\cref{fig:x1}). Placing every job on the device with the best
average fidelity sends all of them to one QPU: mean waiting time of about 2.9--3.0\,ks at every load,
and 40\% of the jobs still finish below $\tau$. A quality-blind shortest-queue balancer barely waits but
violates the requirement for 64--67\% of jobs. The estimate filter followed by earliest expected
completion has no violations and a mean wait of 0.25--8\,s as the load grows from 0.25 to 1.0, with a
decision latency of 4\,ms per job; the same rule on the transpiled circuit has 4.8\% violations,
54--92\,s mean wait and 125\,ms per decision. The price is the filter's conservatism in simulation: it
rejects 18.5\% of jobs that some device could have served, against 3\% for the exact check.

\paragraph{Cost and quota.} With the usage model and illustrative list prices (\cref{tab:cost}),
raw-fidelity placement always chooses the most expensive device and spends USD 12.2k on the 102 E3 jobs
while missing $\tau$ for 26--79 of them. The estimate filter runs fewer jobs, because it rejects those no
device is predicted to serve, but meets $\tau$ for 3--10 jobs per 100 USD against 0.2--0.6 for
raw-fidelity placement. Under a quota of 25\% of the raw-fidelity spend, raw-fidelity placement has 74
jobs blocked and the estimate filter none.

\begin{table}[t]
\centering\small
\caption{Cost and quota under the E3 policies (102 jobs, eight simulated devices), with illustrative
list prices (not quotes). Jobs meeting $\tau$ per 100 USD measure how much useful work the budget buys;
the last column counts jobs blocked when the budget is 25\% of what raw-fidelity placement spends at the
same $\tau$. Ranges are over $\tau=0.5,\dots,0.9$.}\label{tab:cost}
\begin{tabular}{@{}lcccc@{}}
\toprule
Policy & Cost (USD) & Violations & Met $\tau$ per 100 USD & Blocked, 25\% quota\\
\midrule
Highest raw fidelity & 12{,}210 & 26--79 & 0.2--0.6 & 74\\
Lowest cost & 207 & 32--81 & 10--34 & 0\\
Estimate filter (ours) & 136--2{,}184 & 0--1 & 3.2--9.6 & 0\\
Transpile every candidate & 260--4{,}323 & 0--2 & 1.8--8.1 & 0--23\\
\bottomrule
\end{tabular}
\end{table}

\begin{figure}[t]
  \centering
  \includegraphics[width=\linewidth]{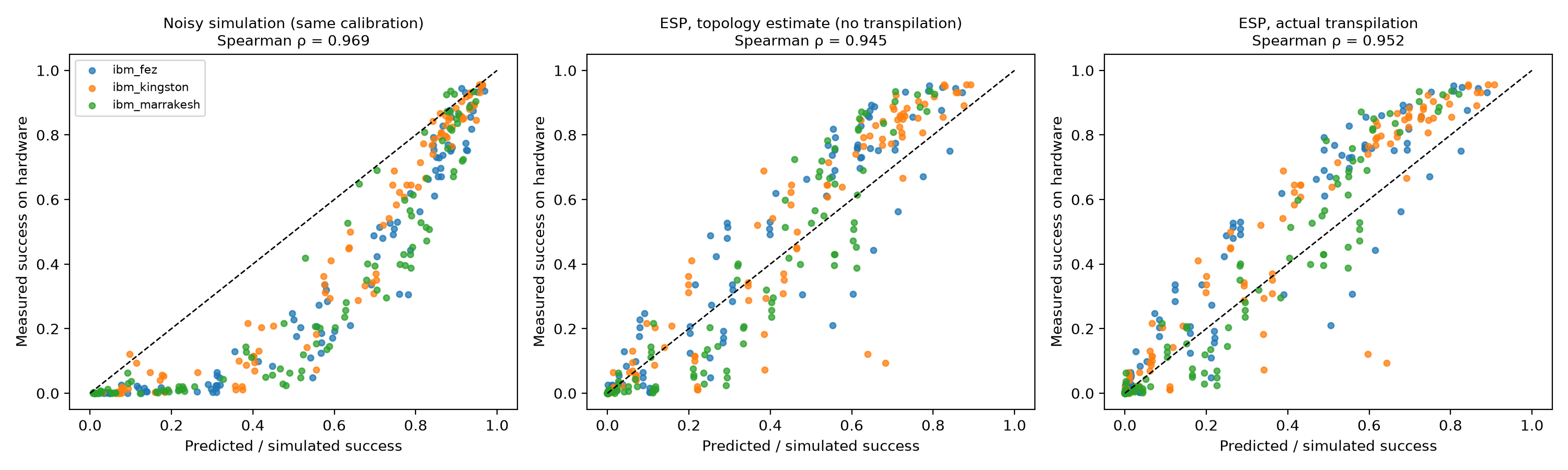}
  \caption{E6: predicted or simulated success (horizontal) vs.\ measured success (vertical) on three IBM
Heron devices, one point per (circuit, device), coloured by device; dashed line is agreement. Left to
right: noisy simulation with the day's calibration, ESP from the topology estimate, and ESP from the
transpiled circuit. Simulation sits systematically high; the two ESP panels rank the circuits similarly.}
  \label{fig:e6}
\end{figure}

\paragraph{E6: hardware validation.} On the three IBM Heron devices (\cref{tab:e6},
\cref{fig:e6}), the transpilation-free estimate ranks the measured success as well as the estimate
from the actually transpiled circuit, with the same mean absolute error, and far better than raw
device fidelity. A noisy simulation with the calibration of the day ranks slightly better but
overestimates hardware success by about $0.17$, consistent with noise the model omits (crosstalk,
drift). In the earlier 28-circuit run the estimate was slightly behind ($\rho=0.91$ vs.\ $0.93$).
About 30\% of the executions, mostly deep 12--14-qubit circuits, have a measured success below
$0.05$, so part of the ranking separates feasible from infeasible circuits.
\texttt{ibm\_kingston} was clearly the best device (mean success $0.45$, against $0.38$ and $0.31$),
and all device-level predictors pick it, with mean regret $0.008$. Between the two near-identical
devices the result is negative: the ESP estimates pick the better of \texttt{ibm\_fez} and
\texttt{ibm\_marrakesh} for only 48\% of circuits (mean regret $0.045$; ties count as hits), while raw
device fidelity picks it for 75\% (regret $0.015$). Device-mean error rates do not separate devices of the same
family. A layout-matched diagnostic confirms where the information is: using the circuits as actually
transpiled and the per-qubit and per-coupler calibration each device published just before the run
(retrieved for that date), an ESP that multiplies the error of every gate on its physical qubits and the
readout error of every measured qubit picks the better of the two for 77\% of circuits (regret $0.013$)
and ranks all executions with $\rho=0.967$. It needs the transpiled circuit, so it is an upper bound for
diagnosis rather than a predictor for the fast stage.

\begin{table}[t]
\centering
\caption{E6: predictors of the success probability measured on three IBM Heron devices (102 mirror circuits
per device, 306 executions, 2000 shots each). Spearman $\rho$ measures ranking against the measured
success and MAE the mean absolute error. The estimate without transpilation ranks as well as the estimate
from the transpiled circuit; raw device fidelity does not rank circuits.}
\label{tab:e6}
\begin{tabular}{lcc}
\toprule
Predictor & Spearman $\rho$ & MAE\\
\midrule
Noisy simulation (same calibration) & 0.97 & 0.17\\
Raw device fidelity & 0.17 & --\\
\textbf{ESP, topology estimate (no transpilation)} & 0.95 & \textbf{0.10}\\
ESP, actual transpilation & 0.95 & 0.10\\
\bottomrule
\end{tabular}
\end{table}

\begin{figure}[t]
  \centering
  \includegraphics[width=0.85\linewidth]{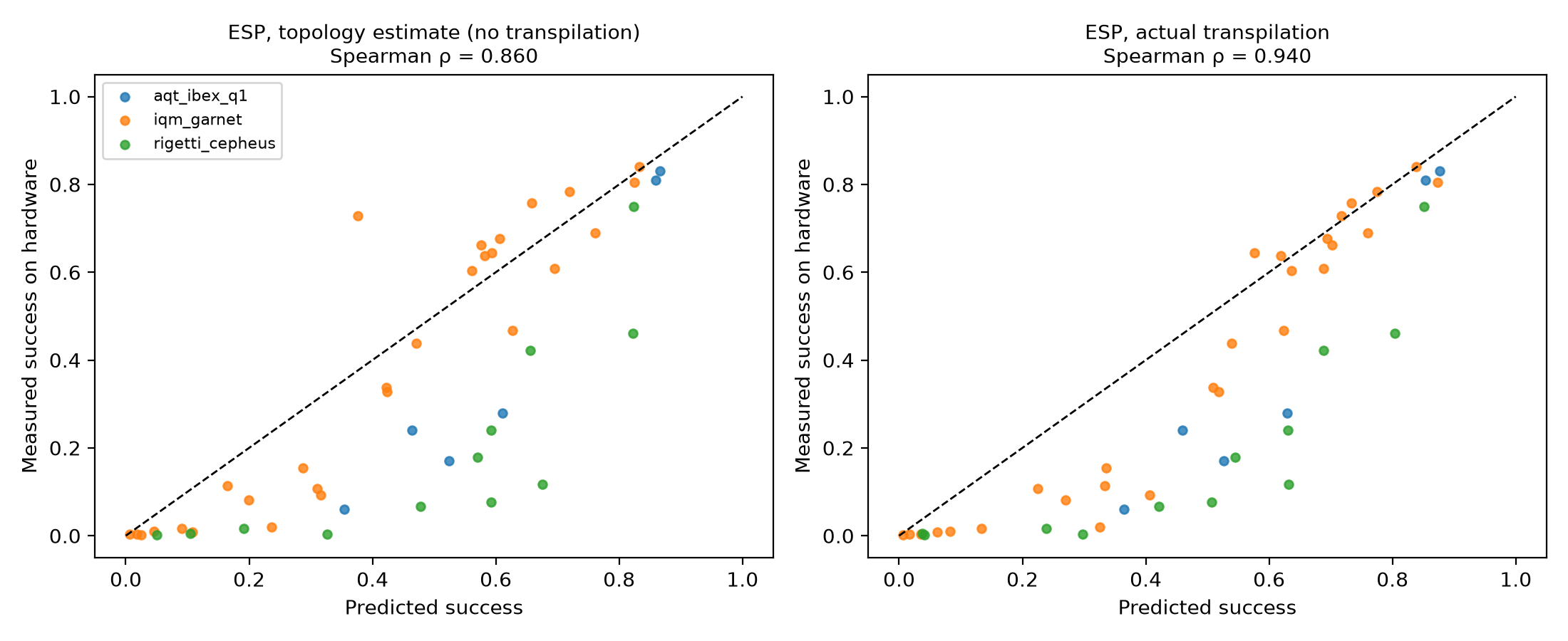}
  \caption{E7: predicted vs.\ measured success on IQM Garnet, Rigetti Cepheus-1-108Q and AQT IBEX Q1, one
point per circuit, coloured by device. Left: ESP from the topology estimate; right: ESP from the
transpiled circuit. Points below the diagonal on Rigetti and AQT show that published error rates
overstate those devices; the ordering of circuits still largely holds.}
  \label{fig:e7}
\end{figure}

\paragraph{E7: other vendors.} Within each device (\cref{tab:e7}, \cref{fig:e7}), the estimate
ranks circuits well, slightly below the estimate from the transpiled circuit: $\rho=0.93$ vs.\
$0.98$ on IQM Garnet and $0.90$ vs.\ $0.94$ on Rigetti; on AQT both reach $0.94$ ($n=6$). Its
predicted two-qubit gate counts follow the actual ones closely ($\rho=0.997$ and $0.993$; on the
all-to-all AQT device no SWAPs are needed and the counts match exactly). On Rigetti and AQT, where
we only have published error figures, measured success is well below both predictions (mean
absolute error about $0.3$ and $0.2$; on AQT, $0.06$--$0.28$ against $0.35$--$0.63$ on the four
deeper circuits, as predicted by either estimate): the absolute prediction is poor, although the ranking holds.

For target selection we combine these runs with the 28-circuit IBM run, giving six devices of four
vendors. \texttt{ibm\_kingston} is the best device for most circuits. Over the five superconducting
devices, the estimate always picks an IBM device: mean regret $0.010$ and the best device for 89\%
of circuits, the same as always choosing \texttt{ibm\_kingston}, while the estimate from the
transpiled circuit picks IQM Garnet for 5 circuits on which it is clearly worse (regret $0.049$,
71\%). Adding AQT changes this. With its published figures, AQT has the highest predicted success
on 5 of its 6 circuits for the estimate and on all 6 for the estimate from the transpiled circuit,
but it is never the best device in the measurements (for example qft\_6: $0.06$ on AQT against
$0.47$ on \texttt{ibm\_fez}). Over six devices the mean regret rises to $0.051$ (top-1 75\%)
for the estimate and $0.096$ (57\%) for the estimate from the transpiled circuit, worse than always
choosing \texttt{ibm\_kingston} ($0.010$). The comparison mixes days and shot counts (500 to 4000
shots), and AQT ran only 6 circuits, so its selection results are not directly comparable with the
28-circuit devices. Both predictors fail here in the same way, so the cause
is the error rates, not the topology estimate: cross-vendor selection needs error rates that are
comparable across vendors, such as per-job calibration or the scheduler's own measurements. The
comparison is not fully controlled: the IBM runs are from an earlier day, and shot counts differ.

We test the second option with one factor $\gamma_d$ per device: the predicted success is
$\mathrm{ESP}^{\gamma_d}$, which approximately scales all of the device's error rates by $\gamma_d$, fitted by least
squares on circuits other than the one being predicted. Fitted on all circuits (topology estimate),
$\gamma$ is $2.2$ for AQT and $3.1$ for Rigetti, against $0.7$--$0.8$ for the IBM devices and $1.1$ for
IQM Garnet. Leave-one-circuit-out, the estimate no longer picks AQT: over six devices the mean
regret is $0.010$ and the best device is picked for 89\% of circuits, with both predictors, equal to
always choosing \texttt{ibm\_kingston} (which is what the learned rates pick); the mean absolute
error on AQT and Rigetti drops from $0.21$ and $0.30$ to $0.08$ for both. A scheduler would learn
from fewer jobs. With $k$ random probe circuits per device (200 repetitions), AQT is avoided already
at $k=1$, but $\gamma$ is noisy (for \texttt{ibm\_kingston} at $k=3$, the 10th--90th percentile range
is $0.1$--$3.3$) and the choice among the three similar IBM devices suffers: the mean regret of the
estimate is $0.061$ at $k=1$ and $0.043$ at $k=5$, with top-1 45\% and 59\% of
circuits. Learned rates thus fix the bias between vendors, and separating similar devices needs
more measurements per device.

\begin{table}[t]
\centering
\caption{E7: hardware validation on three non-IBM devices. Upper rows: Spearman $\rho$ / MAE of each predictor
against the measured success per device (Rigetti and AQT circuits were submitted verbatim). Lower rows:
choosing a device per circuit among five superconducting devices (three IBM, IQM, Rigetti) and among all
six, reported as mean regret and top-1 (share of circuits for which the best device was chosen).}
\label{tab:e7}
\resizebox{\linewidth}{!}{\begin{tabular}{lcc}
\toprule
 & ESP, topology estimate & ESP, actual transpilation\\
\midrule
IQM Garnet (28 circuits) & 0.93 / 0.09 & 0.98 / 0.10\\
Rigetti Cepheus-1-108Q (12 circuits, verbatim) & 0.90 / 0.30 & 0.94 / 0.28\\
AQT IBEX Q1 (6 circuits, verbatim) & 0.94 / 0.21 & 0.94 / 0.22\\
\midrule
Selection over 5 devices: mean regret & \textbf{0.010} & 0.049\\
Selection over 5 devices: top-1 & \textbf{89\%} & 71\%\\
Selection over 6 devices (with AQT): mean regret & \textbf{0.051} & 0.096\\
Selection over 6 devices (with AQT): top-1 & \textbf{75\%} & 57\%\\
Selection over 6 devices, learned $\gamma$ (leave-one-out): mean regret & 0.010 & 0.010\\
Selection over 6 devices, learned $\gamma$ (5 probes): mean regret & 0.043 & 0.037\\
Always \texttt{ibm\_kingston}: mean regret & \multicolumn{2}{c}{0.010}\\
\bottomrule
\end{tabular}}
\end{table}

\paragraph{E12: controlled cross-vendor run.} With identical circuits, shots and day
(\cref{tab:e12}), the estimate ranks circuits within each IBM device with $\rho=0.82$--$0.91$ and on IQM
Garnet with $0.90$, below the estimate from the transpiled circuit ($0.87$--$0.96$), as in E6 and E7.
Choosing among the three IBM devices and IQM, where published error rates match measured behaviour, the
estimate loses $0.005$ of success on average (95\% interval $[0.000, 0.012]$) and picks the best device
for 82\% of the 28 circuits, against $0.025$ and 71\% with the transpiled circuit. On Rigetti the same
day the published medians describe the device poorly ($\rho=0.57$, mean absolute error $0.32$, against
$0.90$ and $0.30$ in E7); selection over five devices on its 12 circuits still loses only $0.007$
($[0.000, 0.020]$) because the estimate does not choose Rigetti. Cross-vendor selection therefore works
when error rates are comparable across vendors and fails to the extent that they are not.

\begin{table}[t]
\centering\small
\caption{E12: controlled cross-vendor run on 20 Nov 2026 (same circuits, 1000 shots, same day). Upper
rows: Spearman $\rho$ against measured success within each device, for ESP from the topology estimate and
from the transpiled circuit. Lower rows: device selection per circuit, mean regret with 95\% bootstrap
interval / top-1. Selection works across IBM and IQM, whose error rates are comparable; Rigetti's
published medians overstate the device.}\label{tab:e12}
\begin{tabular}{@{}lccc@{}}
\toprule
& Circuits & ESP, topology estimate & ESP, transpiled circuit\\
\midrule
\texttt{ibm\_fez} & 28 & 0.82 & 0.87\\
\texttt{ibm\_marrakesh} & 28 & 0.83 & 0.88\\
\texttt{ibm\_kingston} & 28 & 0.91 & 0.95\\
IQM Garnet & 28 & 0.90 & 0.96\\
Rigetti Cepheus-1-108Q & 12 & 0.57 & 0.59\\
\midrule
Selection, IBM $\times3$ + IQM & 28 & \textbf{0.005} $[0.000, 0.012]$ / 82\% & 0.025 $[0.005, 0.049]$ / 71\%\\
Selection, + Rigetti & 12 & \textbf{0.007} $[0.000, 0.020]$ / 83\% & 0.032 $[0.000, 0.071]$ / 67\%\\
\bottomrule
\end{tabular}
\end{table}

\paragraph{E8: training-free estimate vs.\ a learned predictor.} In the spirit of
Fulginiti et al.~\cite{fulginiti2026}, we trained gradient-boosted models on pre-transpilation
circuit features plus device calibration and topology features (target: success or its
logarithm), and a variant that learns a correction factor on top of our estimate. All predictors
are scored on the same out-of-fold rows; strict splits hold out both the device and the circuits
(\cref{tab:e8}). When the target device is in the training data, the learned model is more
accurate, mainly because it learns each device's offset. On an unseen simulated device it is slightly better than our
estimate (MAE $0.08$ vs.\ $0.13$, regret $0.013$ vs.\ $0.016$). On an unseen IBM hardware device, on an
unseen topology, or when trained on simulation and applied to IBM hardware, it ties or loses and chooses
devices worse (on unseen IBM devices top-1 61\% vs.\ 86\%). The exception is the cross-vendor
choice among IQM, Rigetti and AQT, where the learned model trained on simulation picks better
devices although its absolute error is larger; our estimate inherits the published Rigetti and AQT
error rates, which overstate those devices (E7). With 18 circuits this is suggestive only. A
multilayer perceptron collapses when it must extrapolate device features ($\rho=0.04$ on an unseen
topology). The training-free estimate is therefore the robust default for a device that has just
been added or recalibrated, and a learned per-device correction on top of it improves absolute
accuracy (MAE $0.10\to0.07$ on \texttt{ibm\_kingston}, with the device and the circuits held out;
the correction is learned on the other devices). The three
IBM devices are near-identical, so device features hardly vary; we did not run the authors' graph neural network or
MQT Predictor themselves.

\begin{table}[t]
\centering\small
\caption{E8: our training-free estimate vs.\ a learned gradient-boosted predictor, evaluated on the same
out-of-fold rows. Strict splits hold out both the device and the circuits. The learned model is more
accurate on devices seen in training; on an unseen IBM device or topology our estimate ranks and selects
better. The regret of the device choice uses one model for all devices where available; $^*$combines
per-device models, which is pessimistic for the learned model.}\label{tab:e8}
\resizebox{\linewidth}{!}{\begin{tabular}{@{}lccc@{}}
\toprule
Evaluation & Ours $\rho$ / MAE & Learned $\rho$ / MAE & Regret ours vs.\ learned\\
\midrule
Simulation, devices seen & 0.95 / 0.13 & 0.99 / 0.03 & 0.016 vs.\ 0.011\\
Simulation, unseen device (strict) & 0.95 / 0.13 & 0.94 / 0.08 & 0.016 vs.\ 0.013\\
Simulation, unseen topology & 0.95 / 0.13 & 0.81 / 0.20 & 0.016 vs.\ 0.158$^*$\\
IBM hardware, devices seen & 0.95 / 0.10 & 0.95 / 0.06 & 0.008 vs.\ 0.011\\
IBM hardware, unseen device (strict) & 0.95 / 0.10 & 0.92 / 0.11 & \textbf{0.008} vs.\ 0.031\\
Simulation $\to$ IBM hardware & 0.95 / 0.10 & 0.93 / 0.11 & 0.008 vs.\ 0.038\\
Simulation $\to$ IQM/Rigetti/AQT (46 rows) & 0.86 / 0.16 & 0.91 / 0.27 & 0.092 vs.\ \textbf{0.010}\\
\bottomrule
\end{tabular}}
\end{table}

\begin{figure}[t]
  \centering
  \includegraphics[width=\linewidth]{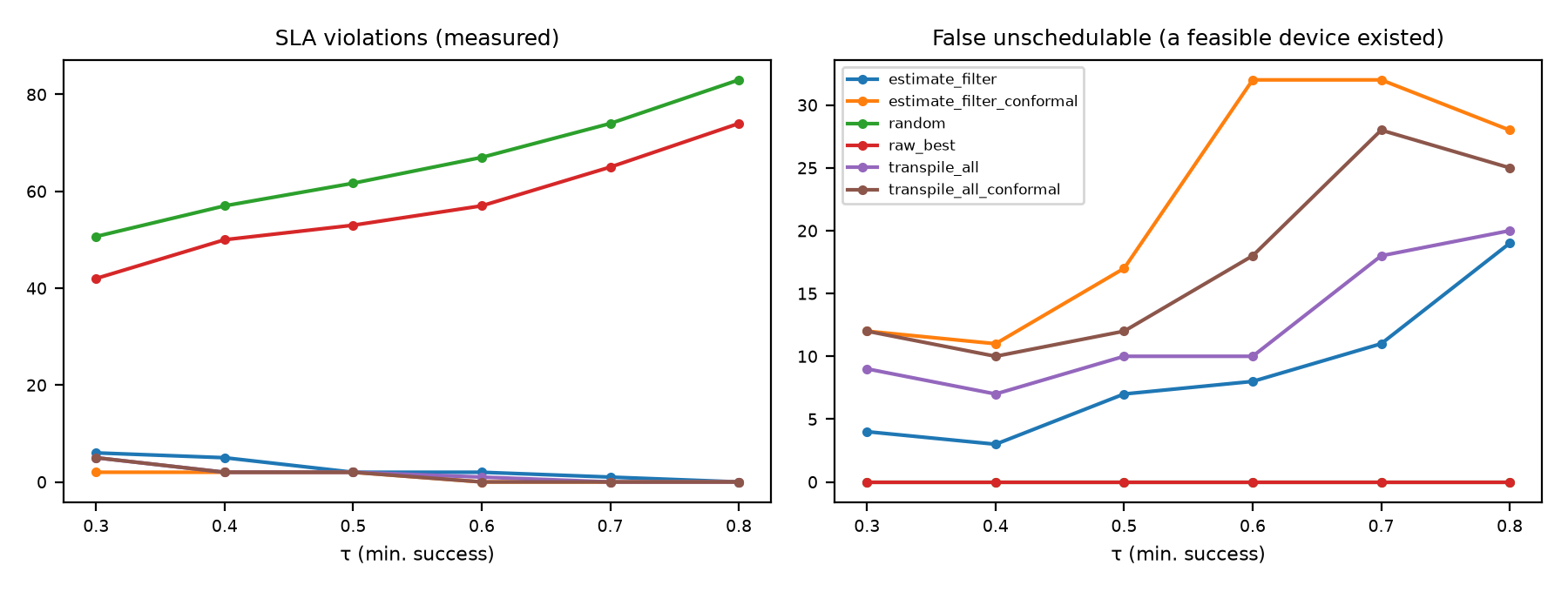}
  \caption{E9 on the three IBM Heron devices (102 jobs), replayed on measured outcomes: SLA violations
(left) and false rejections (right) vs.\ required success $\tau$. Raw-fidelity and random placement put
many jobs below $\tau$ (42--74 and 51--83 of 102); the estimate filter, with or without a conformal
margin, removes almost all of those violations.}
  \label{fig:e9}
\end{figure}

\paragraph{E9: placement policies on measured outcomes.} In E6 every circuit ran on all three IBM
devices, and in E7 the 28 circuits ran on up to six devices of four vendors, so any placement can
be evaluated after the fact on the success the job would have measured on each candidate,
with no new QPU time. A violation is a job placed on a device where its measured success is below
$\tau$; a false rejection is a job declared unschedulable although some device met $\tau$
(\cref{tab:e9}, \cref{fig:e9}). On measured hardware the hard constraint does what E3 showed in
simulation: on IBM it removes almost all violations of raw-fidelity placement (and about half or
more across vendors) and matches transpiling every candidate, with fewer false rejections on IBM. On hardware the estimate is not uniformly
conservative: on IBM its error is centred near zero (mean $-0.03$; it over-predicts 41\% of
pairs), and with published figures it over-predicts by $0.21$ on AQT and $0.30$ on Rigetti. A
split-conformal margin on the over-prediction, calibrated on held-out circuits, reaches the nominal
coverage (0.91 on IBM, 0.92 across vendors, for $1-\delta=0.9$) and cuts violations to 0--2, at the
price of more false rejections. Calibrated on the other circuit families, coverage falls to
0.87 on IBM: a new family breaks exchangeability, so the guarantee holds only for workloads like
the calibration set.

\begin{table}[t]
\centering\small
\caption{E9: placement policies replayed on measured outcomes, over requirements $\tau=0.3,\dots,0.8$: SLA
violations (jobs placed on a device where they measured below $\tau$) and false rejections (jobs rejected
although some device met $\tau$). The conformal margin uses $\delta=0.1$, calibrated on held-out circuits.
Filtering with the estimate removes almost all violations of raw-fidelity placement on IBM, at the cost
of more rejections.}\label{tab:e9}
\begin{tabular}{@{}lcccc@{}}
\toprule
& \multicolumn{2}{c}{IBM $\times3$, 102 jobs} & \multicolumn{2}{c}{6 devices, 4 vendors, 28 jobs}\\
\cmidrule(lr){2-3}\cmidrule(l){4-5}
Policy & Violations & False rej. & Violations & False rej.\\
\midrule
Random device (expected) & 51--83 & 0 & 11--22 & 0\\
Highest raw fidelity & 42--74 & 0 & 6--16 & 0\\
Estimate filter (ours) & 0--6 & 3--19 & 0--3 & 0--8\\
Transpile every candidate & 0--5 & 7--20 & 0--6 & 0--8\\
Estimate + conformal margin & 0--2 & 10--32 & 0--1 & 7--15\\
\bottomrule
\end{tabular}
\end{table}

\begin{figure}[t]
  \centering
  \includegraphics[width=\linewidth]{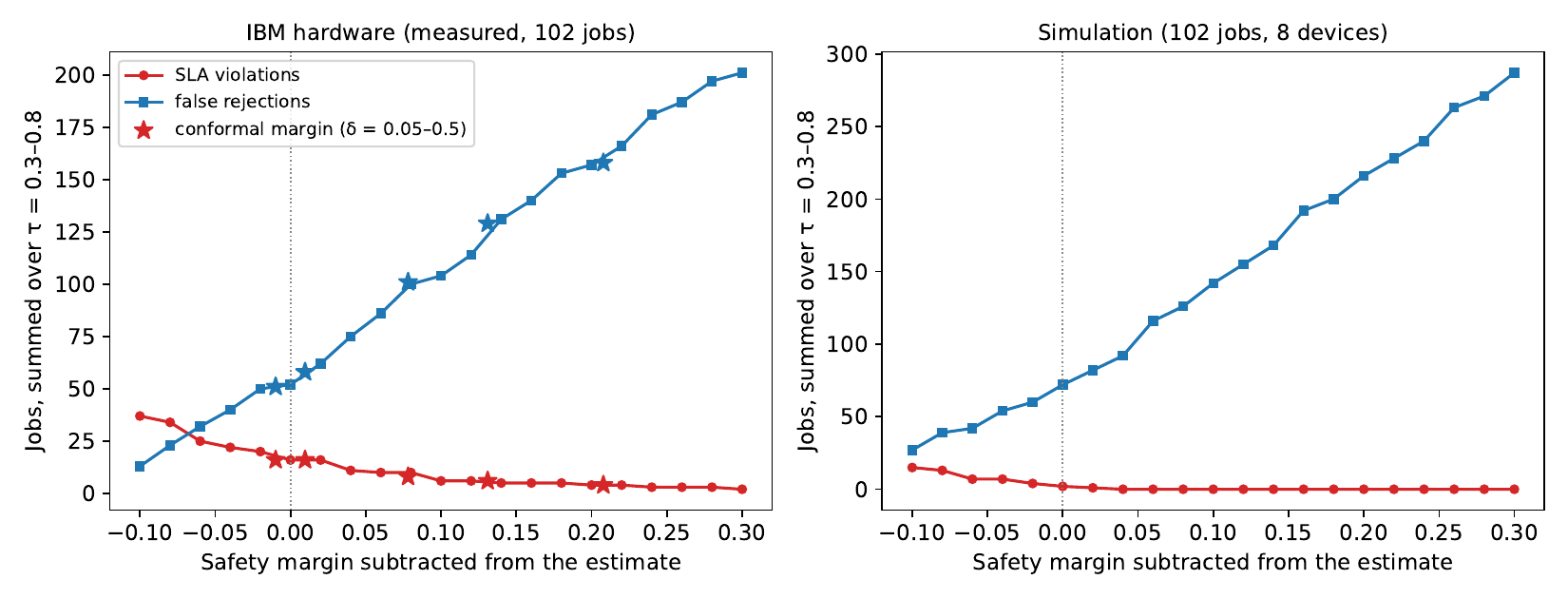}
  \caption{Operating curve of the safety margin subtracted from the estimate before comparing with
$\tau$, on IBM measured outcomes (left) and in simulation (right). Each point sums SLA violations (red)
and false rejections (blue) over $\tau=0.3,\dots,0.8$ and 102 jobs. Stars mark split-conformal margins at
$\delta=0.05$--$0.5$. A negative margin admits more jobs and more violations, a positive one fewer of
both; the conformal margins lie on the same curve.}
  \label{fig:margin}
\end{figure}

The margin is the operator's control over that trade-off (\cref{fig:margin}). Summed over
$\tau=0.3,\dots,0.8$ on IBM, the filter without margin has 16 violations and 52 false rejections; a
margin of $-0.1$ gives 37 and 13, and $+0.1$ gives 6 and 104, against 341 violations for raw-fidelity
placement. The conformal margins at $\delta=0.05$--$0.5$ fall on the same curve, so they choose a point on
it with a stated coverage rather than move it.

\begin{figure}[t]
  \centering
  \includegraphics[width=\linewidth]{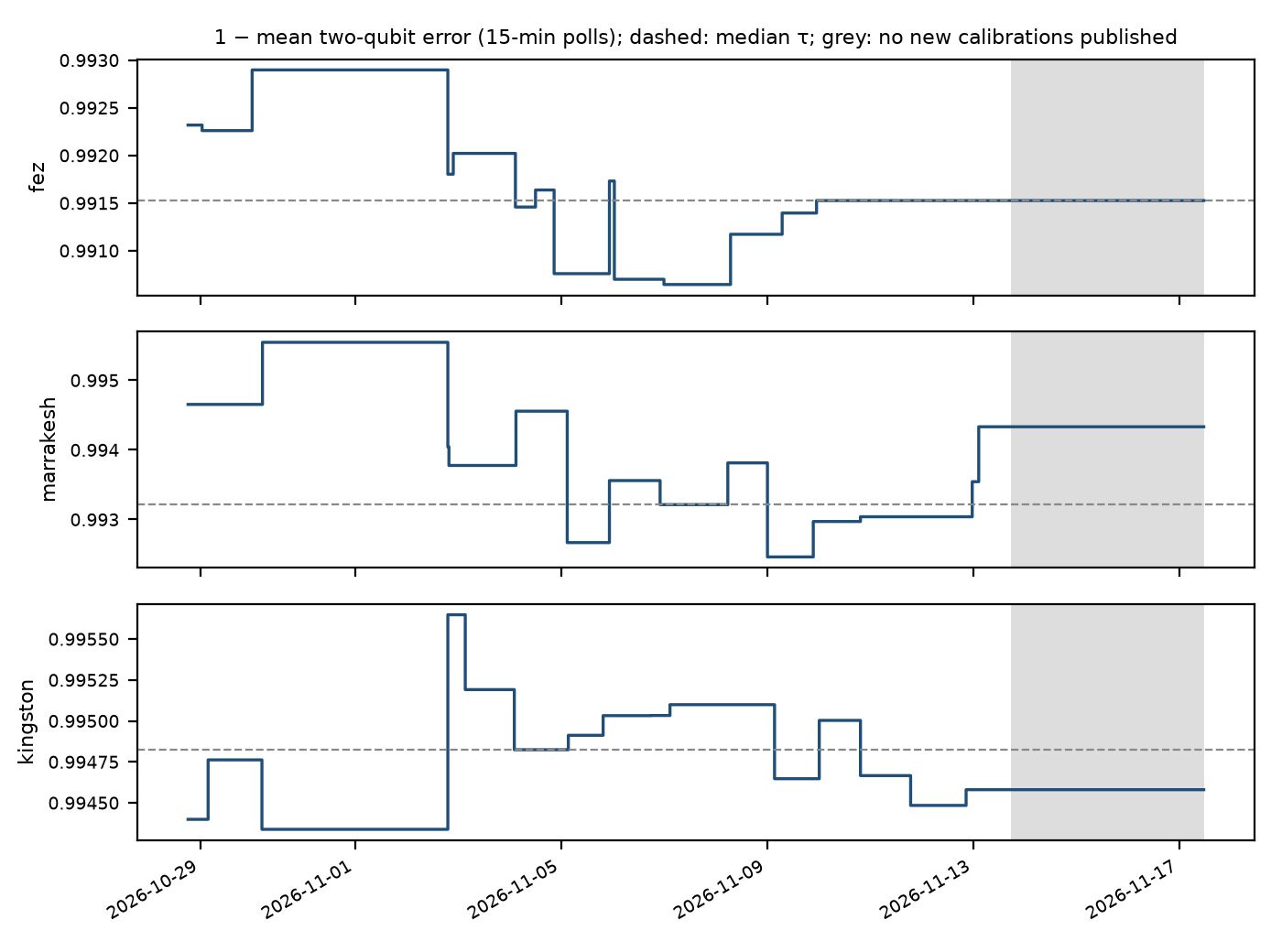}
  \caption{E4: quality of three IBM Heron devices over time, taken as $1-$ mean two-qubit error from
15-minute polls of the published calibration. The dashed line is the per-device median used as $\tau$;
the grey band marks the stretch where published data stopped changing. The series is piecewise constant,
with roughly one jump per day---too sparse for a short-horizon forecast to anticipate.}
  \label{fig:e4}
\end{figure}

\paragraph{E4/E5: forecasting and deferral on real history (negative result).} We polled
\texttt{ibm\_fez}, \texttt{ibm\_marrakesh} and \texttt{ibm\_kingston} every 15 minutes from 28 Oct
to 13 Nov 2026, 11.6 days of coverage after removing collection gaps; grid points more than 30
minutes from a real poll are not scored, and jobs whose deadline window overlaps a gap are
skipped. Between 13 Nov (17:22 UTC) and 18 Nov the published calibration data of the three devices did
not change while they continued to run jobs, so we end the history on 13 Nov. At the published resolution
the mean qubit fidelity varies little (3--6 distinct values per device), so we use quality $=1-$ mean
two-qubit error, which drives ESP. Even this series is piecewise constant, with 10--12 changes per
device in 11.6 days (about one per day; \cref{fig:e4}), although the update timestamp
advances every $\sim$30 minutes. For E4, which asks whether quality windows are reliable, over a 6\,h horizon persistence reaches precision 0.95 and
recall 0.97; the calibration-cycle forecaster's lower bound almost never predicts a window
(precision 1.00, recall 0.01), its mean reaches 0.68/0.93, and Prophet's lower bound 0.87/0.54.
For E5, which asks whether deferral helps under deadlines, for jobs whose device fails $\tau$ at arrival, an oracle that waits for the first good moment
meets the SLA for 6\%, 16\% and 38\% of jobs with 3, 12 and 24\,h deadlines (71, 56 and 55 jobs); our deferral policy
for none. On simulated history that follows the forecaster's own 12\,h cycle, the same code gives
39\% deferral success (oracle 44\%): the mechanism works when the dynamics match the model, and the
model does not describe these devices. Exploiting day-scale headroom would require modelling when
the daily step happens, which about ten events per device cannot fit.

\paragraph{Calibration several days old.} The period without calibration updates allowed a natural
experiment. On 17 Nov we reran the 102 mirror circuits of E6 on the three devices (306 executions,
177\,s of QPU), predicting with the latest published calibration, from 13 Nov. The first
\texttt{ibm\_kingston} job did not complete and was resubmitted. The same circuits measured almost the same as
on 3 Nov ($\rho=0.977$ between the two runs, mean difference $+0.007$). With four-day-old calibration the
estimate ranked measured success as well as with fresh calibration ($\rho=0.951$ vs.\ $0.945$, MAE $0.096$
vs.\ $0.099$) and still chose among the three devices well (regret $0.003$, top-1 84\%), whereas raw device
fidelity, which had picked the best device on 3 Nov, now almost always picked a worse one (regret $0.078$,
top-1 8\%). On these devices, a few days of staleness did not hurt the circuit-specific estimate, while
the device-level ranking it uses as input became less informative. This is one device family over four days.

\paragraph{E10: application circuits.} Mirror circuits have one known output; application circuits
do not. We ran 19 application circuits without mirroring (Bernstein--Vazirani and GHZ with 4--10 qubits, Grover with
3--4, a ripple-carry adder with 6 and 8, QAOA MaxCut on a ring and on random 3-regular graphs, a
hardware-efficient ansatz of the kind used in the variational quantum eigensolver (VQE) with 4--8) on the three IBM Heron devices (2000 shots), Rigetti Cepheus (1000
shots, verbatim) and AQT IBEX Q1 (17 circuits up to 8 qubits, 200 shots, verbatim), 93 executions in total.
Quality is the normalized Hellinger fidelity to the ideal distribution~\cite{lubinski2021}; the IBM runs used
the 13 Nov calibration, the latest published at the time (\cref{tab:e10}). Within each device the estimate ranks circuits as well as the
estimate from the transpiled circuit, as for mirror circuits, but the rankings are weaker. Part of the gap is
the metric: ESP is the probability of an error-free run, while the Hellinger fidelity of a broad output
distribution (QAOA, VQE, Grover) tolerates errors that leave the distribution nearly unchanged, so on IBM ESP
underestimates it by 0.13--0.17. Choosing among the three IBM devices still works: mean regret $0.009$ with the
estimate (top-1 63\%) vs.\ $0.063$ with raw device fidelity (top-1 11\%), the same as with the transpiled
circuit. Across vendors selection fails (regret $0.24$): with published figures the estimate overstates Rigetti
by $0.47$ and AQT by $0.25$. On Rigetti the raw bitstrings show specific qubits or couplers failing (GHZ-4
returns 0000 and 0011, so one pair never entangled), which a device-mean error rate cannot see. Rankings also do not hold within families with broad
output distributions (Spearman $\rho$ of the estimate within QAOA on 3-regular graphs $-0.07$, QAOA on a
ring $-0.18$, VQE $0.11$), probably because the Hellinger fidelity of a broad distribution depends on how
errors reshape it, which ESP does not model. Split by output type, the estimate ranks the IBM executions
with $\rho=0.71$ for circuits with a concentrated output (Bernstein--Vazirani, GHZ, adder, Grover;
transpiled circuit $0.74$) and $0.55$ for broad outputs (QAOA, VQE); pooled over all vendors, broad
outputs fall to $0.07$. Our claims therefore concern success probabilities with a structured target, such as
mirror circuits and concentrated application outputs; E10 is a transfer check rather than a validation
for arbitrary output distributions.

\begin{table}[t]
\centering\small
\caption{E10: 19 application circuits run without mirroring on five devices. Quality is the normalized
Hellinger fidelity to the ideal output distribution; Spearman $\rho$ compares that fidelity to ESP from
the topology estimate and from the transpiled circuit; bias is mean(ESP $-$ measured). Rankings are
weaker than for mirror circuits, and published error rates overstate Rigetti and AQT.}\label{tab:e10}
\begin{tabular}{@{}lccccc@{}}
\toprule
Device & $n$ & $\rho$ estimate & $\rho$ transpiled & MAE estimate & Bias (ESP$-$measured)\\
\midrule
\texttt{ibm\_fez} & 19 & 0.74 & 0.75 & 0.15 & $-0.15$\\
\texttt{ibm\_kingston} & 19 & 0.77 & 0.82 & 0.13 & $-0.13$\\
\texttt{ibm\_marrakesh} & 19 & 0.48 & 0.50 & 0.18 & $-0.17$\\
Rigetti Cepheus & 19 & 0.73 & 0.73 & 0.47 & $+0.47$\\
AQT IBEX Q1 & 17 & 0.72 & 0.71 & 0.27 & $+0.25$\\
\bottomrule
\end{tabular}
\end{table}

\section{Discussion and Limitations}\label{sec:discussion}
The results support the two-stage design, but several limits apply to how far they generalize.
\begin{itemize}
  \item \textbf{Scope of the prototype.} The prototype implements the scheduling phases as a stand-alone
        pipeline over an internal candidate set; it is not deployed as a Kubernetes or Slurm plugin.
        Our hardware claims concern ranking, selection and threshold filtering replayed on measured
        outcomes; absolute predictions are reliable only with device-specific error rates or a
        calibrated margin.
  \item \textbf{Simulation labels.} The simulation labels come from noisy simulation based on
        calibration snapshots, without crosstalk, drift within a job or leakage. On hardware the
        simulation overestimated success by about 0.17.
  \item \textbf{Hardware coverage.} The hardware validation covers three superconducting vendors and
        one trapped-ion device, but no neutral-atom device, and estimates based on device-mean error
        rates cannot separate near-identical devices.
  \item \textbf{Comparable error rates across vendors.} For Rigetti and AQT we had only published error
        figures. Absolute predictions were poor, and both the estimate and a learned model inherit
        whatever bias those figures carry. Scaling each device's error rates by a factor learned from
        its measured circuits removed the bias, but with five probe circuits or fewer per device the
        factor is too noisy to separate similar devices. In the controlled run (E12), selection works
        between IBM and IQM, whose error rates match measured behaviour, and Rigetti's medians described
        the device even worse than in E7, so day-to-day variation adds to the bias.
  \item \textbf{Provider-side compilation.} Benchmarks that rely on circuit structure, such as mirror
        circuits, require submitting the circuit verbatim, with the provider's own compiler disabled.
  \item \textbf{Per-qubit error rates.} Per-qubit rates performed worse than device means in
        simulation, while the layout-matched diagnostic separates near-identical devices on hardware
        (77\% vs.\ 48\%). Both point to the same next step: an estimate that predicts the compiler's
        placement on good qubits without transpiling.
  \item \textbf{Contention model.} E11 uses simulated devices, first-in-first-out queues without
        preemption or provider-side batching, and service times from the usage model; it shows the
        qualitative behaviour of the policies, not the queueing of a specific provider.
  \item \textbf{Conservatism of the filter.} In simulation ESP underestimates success; on hardware its
        error is centred near zero on IBM devices and positive with published figures. A conformal
        margin gives nominal coverage on held-out circuits but not on held-out circuit families.
  \item \textbf{Calibration dynamics.} On the IBM Heron devices, device-mean quality changes about once a
        day, the update timestamp reflects partial updates rather than a fixed cycle, and the published
        data can stay unchanged for several days, so freshness should be judged on the calibration
        timestamp rather than on the time of the poll. In our rerun, calibration four days old did not
        degrade the circuit-specific estimate but made raw device fidelity choose the wrong device; how
        long published calibration stays usable depends on the device and the workload. The history
        covers 11.6 days and three devices of one vendor.
  \item \textbf{Calibration at run time.} Calibration is moving into run time: pulse-level frameworks
        such as CUDA-Q Pulse~\cite{cudaqpulse} run relaxation-time ($T_1$), Ramsey and gate calibrations while the QPU
        executes, and providers are adding calibration within sessions. A scheduler that relies only on
        published telemetry will see an increasingly outdated picture; consuming measurements taken in
        the session, or requesting a calibration before a critical job, is more promising than
        forecasting the next published calibration.
  \item \textbf{Prices} are illustrative list prices.
\end{itemize}

\section{Conclusion}\label{sec:conclusion}
The goal of this work was to let a cluster scheduler choose the QPU for a job automatically, so that the
job meets its required success probability, within the latency of a scheduling decision. On a single
vendor the goal is met. The estimate runs in milliseconds without compiling, places jobs as well as
compiling for every candidate at about $25\times$ lower latency, ranks measured success on IBM hardware
as well as the estimate computed after compilation, and reduces requirement violations from tens of jobs
to a few compared with picking the device with the best average fidelity. Under simulated contention it keeps
violations at zero without concentrating jobs on one device. Across vendors the goal is met only in
part: selection works when error rates are comparable, as between IBM and IQM in a controlled same-day
run, and fails when published figures overstate a device.

The paper contributes four things that, to our knowledge, have not been combined before. First, it
formulates device choice as a scheduling decision with the user's success requirement as a hard
constraint, rather than as a ranking left to the user. Second, it provides a training-free estimate of
post-compilation success that needs only the circuit graph, the device connectivity and published
calibration, so a new device can be scheduled as soon as it publishes calibration. Third, it gives a
two-stage protocol (estimate, reserve, compile, re-check, bind or roll back) that maps onto the filter
and bind phases of the Kubernetes Scheduling Framework and onto Slurm with QDMI and QRMI; these
integrations already allocate and account for QPU time, and the protocol adds the choice of device.
Fourth, it validates the approach on six devices from four vendors, including a controlled cross-vendor
run, and reports negative results: device-mean error rates do not separate near-identical devices,
published figures can mislead cross-vendor choice, and forecast-based deferral did not help.

The main open problem is comparable error rates across vendors. Next steps are longer multi-vendor
calibration histories, learning error rates from few jobs, an estimate that predicts the compiler's
qubit placement without compiling, and more trapped-ion and neutral-atom devices.

\section*{Acknowledgments}
We thank the colleagues who created accounts on the Open Quantum and IQM Resonance platforms at our
invitation; the credits granted for those sign-ups made the multi-vendor hardware runs (E7, E10 and E12)
possible.

\appendix
\section{Confidence intervals for the main claims}\label{app:ci}
\begin{table}[h]
\centering\small
\caption{95\% paired bootstrap intervals over circuits ($B=2000$) for the main ranking and selection
results. Point estimates are those reported in the text. Overlapping intervals for the topology estimate
and the transpiled circuit support ``as well as''; the wide intervals of the six-device selection (E7) are
why it is presented as a case study.}
\begin{tabular}{@{}llcc@{}}
\toprule
Experiment & Metric & ESP, topology estimate & ESP, transpiled circuit\\
\midrule
E2 (simulation) & Spearman $\rho$ & 0.949 $[0.930, 0.962]$ & 0.942 $[0.929, 0.952]$\\
E2 (simulation) & mean regret & 0.016 $[0.012, 0.020]$ & 0.015 $[0.011, 0.018]$\\
E2 (simulation) & raw fidelity, $\rho$ & \multicolumn{2}{c}{0.371 $[0.335, 0.420]$}\\
E6 (IBM) & Spearman $\rho$ & 0.945 $[0.919, 0.961]$ & 0.952 $[0.928, 0.964]$\\
E6 (IBM) & mean regret & 0.008 $[0.000, 0.020]$ & 0.008 $[0.000, 0.020]$\\
E7 (5 devices) & mean regret & 0.010 $[0.000, 0.027]$ & 0.049 $[0.018, 0.086]$\\
E7 (6 devices) & mean regret & 0.051 $[0.014, 0.094]$ & 0.096 $[0.049, 0.157]$\\
E12 (IBM $\times3$ + IQM) & mean regret & 0.005 $[0.000, 0.012]$ & 0.025 $[0.005, 0.049]$\\
\bottomrule
\end{tabular}
\end{table}

\bibliographystyle{plainnat}
\bibliography{refs}

\end{document}